\documentclass[useAMS,usenatbib]{mn2e}
\usepackage{graphics}
\usepackage{graphicx}
\usepackage{epsf}
\usepackage{url}

\newcommand{\bk}{{\bmath k}}

\def\hMpc{{\ }h^{-1}{\,}{\rm Mpc}}
\def\ihMpc{{\ }h{\,}{\rm Mpc}^{-1}}
\def\bssC{\textbf{\textsf{C}}}
\def\bssF{\textbf{\textsf{F}}}
\def\R{\mathcal{R}}
\def\halofit{{\scshape halofit}}
\def\CAMB{{\scshape camb}}
\def\to3{T^{\mathcal{O}(3)}}
\newcommand{\expect}[1]{\left\langle #1 \right\rangle}  

\newcommand{\dlogs}{
  \begin{figure}
    \begin{center}
     \leavevmode
      \epsfxsize=\columnwidth
      \epsfbox{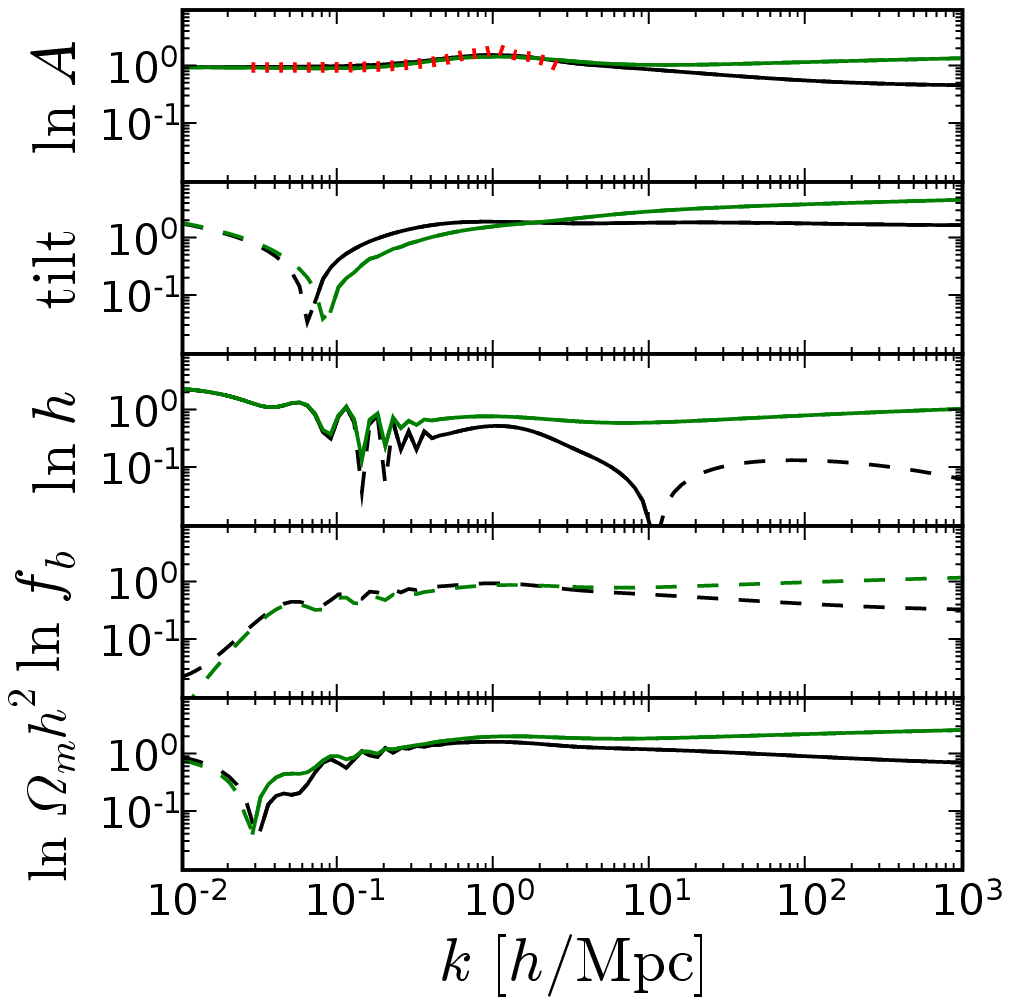}
    \end{center}

    \caption[1]{ \small Parameter derivatives $\partial\ln
    P(k)/\partial \alpha$ (where $\alpha$ is the parameter shown on
    the left), computed using two methods: our halo model code (black
    curves) and the \citet{sea} fitting function (green curves).
    Where the curves are dashed, the derivatives are negative.  The
    red dotted $\partial\ln P(k)/\partial\ln A$ curve was measured
    from 400 $N$-body simulations by RH05.
    \label{dlogs}
    }
  \end{figure}
}
\newcommand{\infodefs}{
  \begin{figure}
    \begin{center}
     \leavevmode
      \epsfxsize=\columnwidth
      \epsfbox{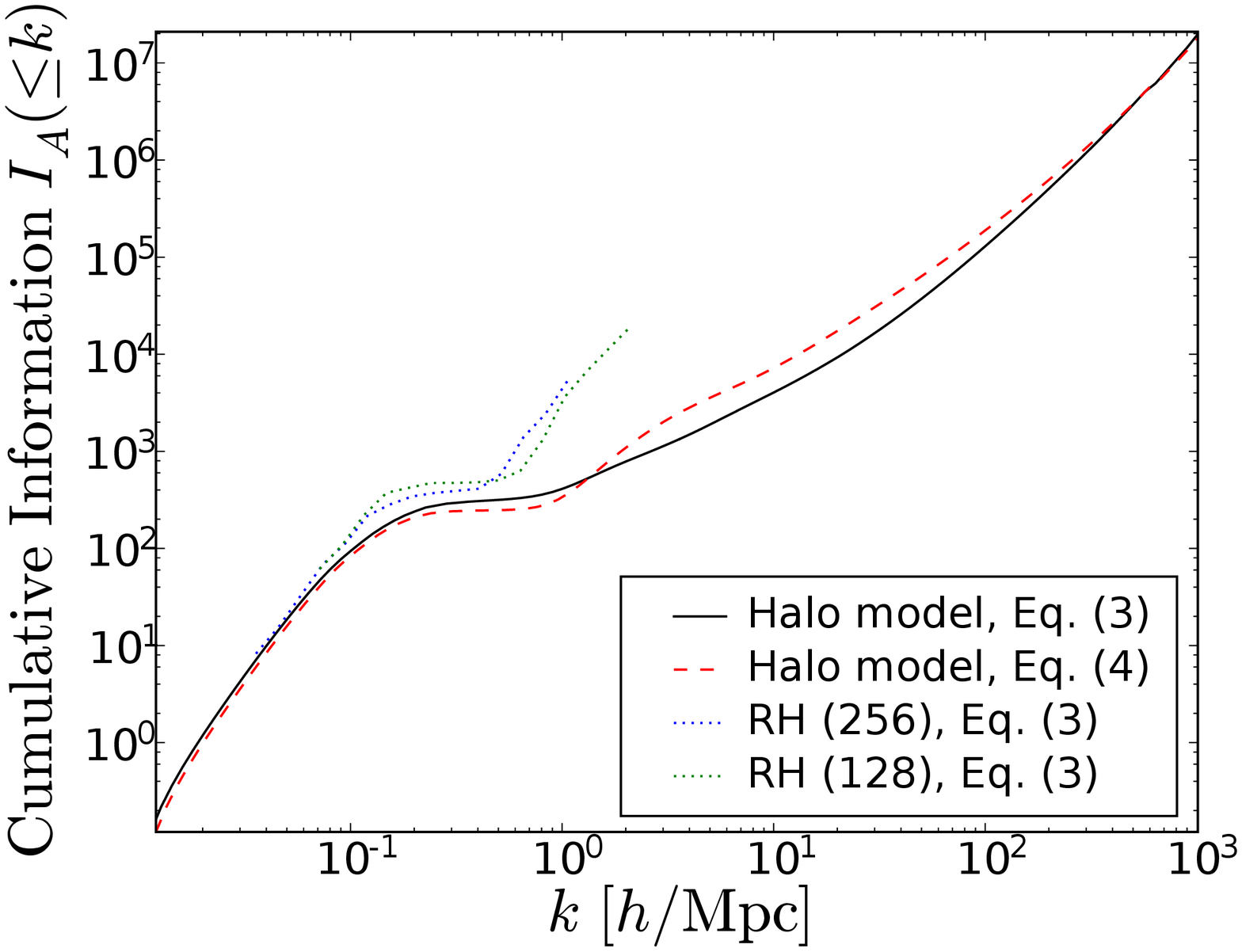}
    \end{center}

    \caption[1]{ \small Different definitions of cumulative
    information, $F_{\ln A,\, \ln A}(\le k)$, about $\ln A$ in the
    dark-matter power spectrum ($I_A(\le k)$ is the notation used in
    Paper I.)  The red, dashed curve uses the incorrect, `old'
    information definition, used in Paper I and RH05, while the black
    curve uses the correct definition, used in this Letter, and RH06.
    The dotted curves show what RH06 found from $N$-body simulations
    of box size 128 and 256$\hMpc$, using the `new' definition.
    \label{infodefs}
    }
  \end{figure}
}

\newcommand{\infomarg}{
  \begin{figure*}
    \begin{minipage}{175mm}
      \begin{center}
	\leavevmode
	\epsfxsize=\columnwidth   
	\epsfbox{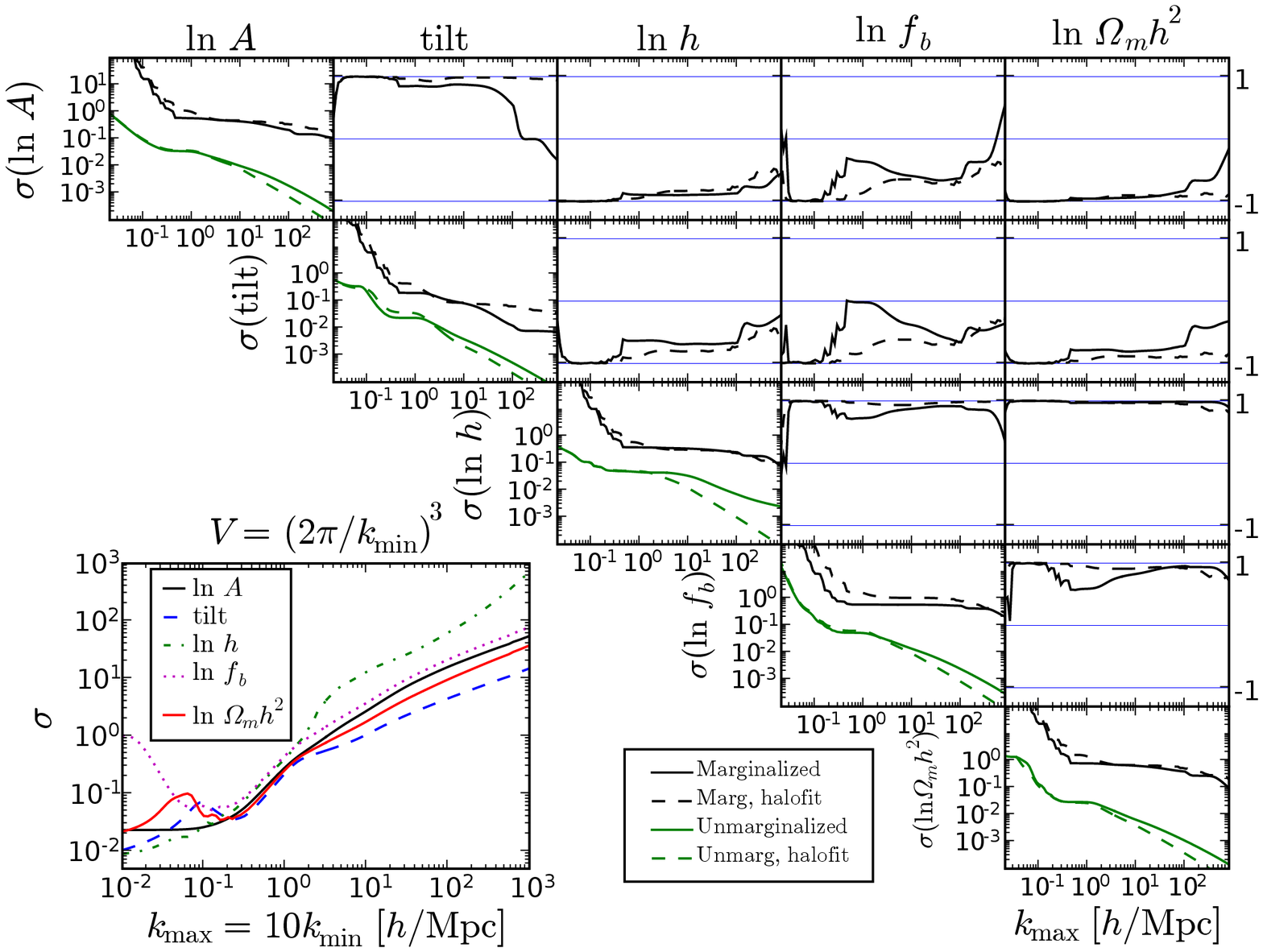}
      \end{center}
      \caption[1]{ \small \textbf{(A, \textit{Upper-right})}.
	One-sigma error-bar half-widths on various cosmological
	parameters as a function of the highest wavelength considered,
	$k_{\rm max}$.  The lowest wavelength is held constant at
	0.02$\ihMpc$, and a volume of $(256\hMpc)^3$ is used.  Plots
	on the diagonal show error-bar half-widths in single
	parameters, both unmarginalized (black) and marginalized over
	all four other parameters (green).  For the solid curves, we
	use our halo model code for parameter derivatives; for the
	dashed, we use \halofit.  Off-diagonal plots show correlation
	coefficients between pairs of parameters, in the marginalized
	covariance matrix $\bssF^{-1}$.
      
	\textbf{(B, \textit{Lower-left})}.  A crude exploration of
	practical issues of survey size.  We show unmarginalized
	one-sigma error-bar half-widths in various cosmological
	parameters, holding the dynamic range of scales used constant,
	at a factor of 10, and using parameter derivatives from our
	halo model code.  The volume of the box changes with $k_{\rm
	max} = 10k_{\rm min}$; we imagine measuring the power spectrum
	in a box of volume $(2\pi/k_{\rm min})^3$.
	\label{infomarg}
      }
    \end{minipage}
    \end{figure*}
}

\newcommand{\masscut}{
  \begin{figure}
      \begin{center}
	\leavevmode
	\epsfxsize=\columnwidth   
	\epsfbox{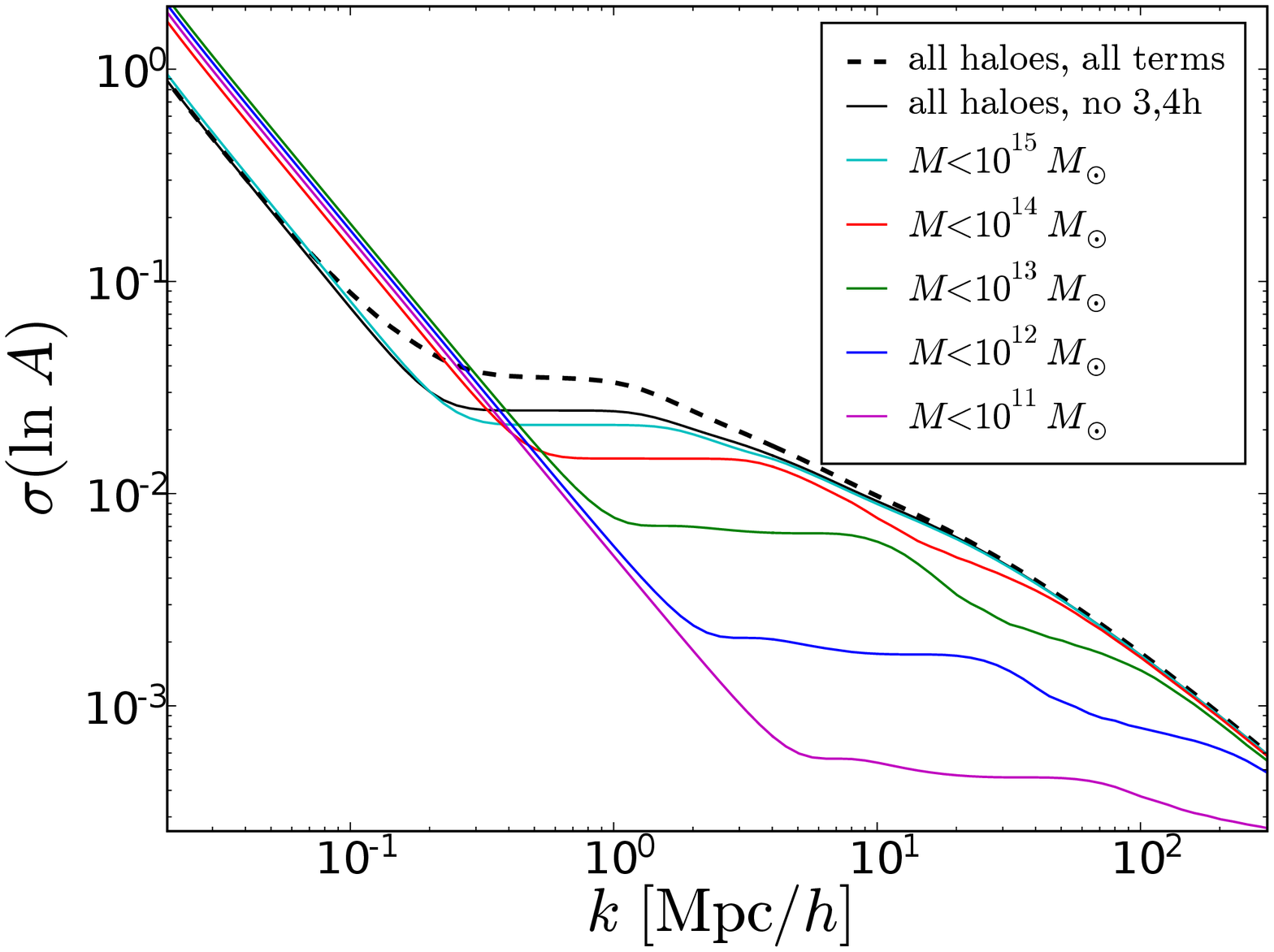}
      \end{center}
      
      \caption[1]{ \small The effect of introducing a mass cut-off in
	the halo mass function on error-bar half-widths on $\ln A$.
	Again, $V=(256\ihMpc)^3$.  The information in the power
	spectrum of matter in low-mass haloes is potentially greater
	than in the full matter distribution.  For the solid curves,
	only the dominant one- and two-halo terms of the halo-model
	trispectrum are included; for the dashed curve, the full
	trispectrum is used, including the three- and four-halo terms.
	\label{masscut}
      }
  \end{figure}
}

\begin{document}
\title[Matter power spectrum information: Multiple parameters]
{Information content in the halo-model dark-matter power spectrum II:
Multiple cosmological parameters}

\author[Mark C.\ Neyrinck and Istv\'{a}n Szapudi]
{Mark C.\ Neyrinck$^1$ and Istv\'{a}n Szapudi$^1$\\
$^{1}$Institute for Astronomy, University of Hawaii, Honolulu, HI 96822, USA\\
  email: {\tt neyrinck@ifa.hawaii.edu}}
\pubyear{2006}

\bibliographystyle{mnras}
	 
\maketitle
  
\begin{abstract}
  We investigate the cosmological Fisher information in the non-linear
  dark-matter power spectrum in the context of the halo model.  We
  find that there is a plateau in information content on translinear
  scales which is generic to all cosmological parameters we tried.
  There is a rise in information on smaller scales, but we find that
  it is quite degenerate among different cosmological parameters
  (except, perhaps, the tilt).  This suggests that it could be
  difficult to constrain cosmological parameters using the non-linear
  regime of the dark-matter power spectrum.  We suggest ways to get
  around this problem, such as removing the largest haloes from
  consideration in survey analysis.
\end{abstract}
\begin {keywords}
  cosmology: theory -- large-scale structure of Universe
\end {keywords}

\section{Introduction}
The distribution of matter in the Universe on large scales contains a
wealth of cosmological information.  Even though galaxy redshift
surveys provide a huge amount of data about galaxy clustering on
non-linear scales, it is unclear how useful these smaller scales are
cosmologically.  Using $N$-body simulations, Rimes \& Hamilton (2005,
RH05; 2006, RH06) investigated the amount of cosmological information,
as a function of scale, in the matter power spectrum $P(k)$.  They
found that information about $\ln A$, the initial amplitude of the
linear power spectrum, is preserved in $P(k)$ on large, linear scales,
and that there is significant information on small scales, but there
is little independent information on translinear scales ($k
\sim0.2-0.8\hMpc$). \citet[][Paper I]{nsr} found that the halo model
also predicts this translinear plateau in the information about $\ln
A$ in $P(k)$.  Paper I showed that the translinear plateau came
largely from cosmic variance in the number of the largest haloes in a
given volume.

In this Letter, we extend the analysis of Paper I, looking at how well
$P(k)$ can constrain multiple cosmological parameters simultaneously.

\section{Method}
The Fisher information matrix $F_{\alpha\beta}$ about parameters
$\alpha$ and $\beta$ given a set of data is defined \citep{fisher,tth}
as the concavity of the natural logarithm of the likelihood function
$\mathcal{L}$, averaged over an ensemble of data predicted by
$\mathcal{L}$;
\begin{equation}
  F_{\alpha\beta} \equiv - \expect{\frac{\partial^2\ln\mathcal{L} ({\rm data} | \alpha, \beta,{\rm priors})}{\partial \alpha \,\partial \beta}}.
\label{iabstract}
\end{equation}

In this Letter, we investigate the (hereafter, implicitly non-linear,
dark-matter) power spectrum $P_i = P(k_i)$ (actually, $\ln P_i$),
measured in bins of wavenumber $k_i$.  The cumulative Fisher information over 
a range of bin indices $i\in\R$ is
\begin{equation}
  F_{\alpha\beta}(\R) = \sum_{i,j\in \R}-\expect{\frac{\partial\ln P_i}{\partial\alpha}
\frac{\partial^2\ln\mathcal{L}}{\partial\ln P_i\partial\ln P_j}\frac{\partial\ln P_j}{\partial\beta}}.
\label{infofishmat}
\end{equation}
To simplify this, we approximate the expectation value of the data
Fisher matrix as $\bssC ^{-1}$, the inverse of the data covariance
matrix $C_{ij} \equiv \expect{\Delta \ln P_i\Delta \ln P_j}$.  This
approximation is good if estimates of $\ln P_i$ have Gaussian
distributions about their expectation values.  This seems to be
adequately so for measurements of the dark-matter power spectrum
(RH06), as the central limit theorem would encourage one to think.  We
denote the data covariance matrix as $\bssC$ (with Fisher matrix
$\bssC^{-1}$), and denote the parameter Fisher matrix as $\bssF$ (with
covariance matrix $\bssF^{-1}$).  Equation (\ref{infofishmat}) becomes
\begin{equation}
  F_{\alpha\beta}(\R) =
  \sum_{i,j\in \R} \frac{\partial\ln P_i}{\partial\alpha}(\bssC_\R^{-1})_{ij} \frac{\partial\ln P_j}{\partial\beta},
  \label{inforange}
\end{equation}
where $\bssC_\R$ is the square submatrix of $\bssC$ with both
indices ranging over $\R$.

The definition in Eqn.\ (\ref{inforange}) is equivalent to the one
used in RH06, except that it bypasses the step of explicit
upper-Cholesky decorrelation.  However, Eqn.\ (\ref{inforange})
differs from the erroneous definition used in RH05 and Paper I, for
which derivative terms were used only on the diagonal.  This previous
definition can be written
\begin{eqnarray}
  F_{\alpha\beta}^{\rm old}(\leq k_{\rm max}) &=&
  \sum_{k_i \le k_{\rm max}} \frac{\partial\ln P_i}{\partial\alpha} \frac{\partial\ln P_i}{\partial\beta}\times\nonumber\\
& &\left[\sum_{l,m\leq i} (\bssC_{\leq i}^{-1})_{lm} - 
  \sum_{l,m\leq i-1} (\bssC_{\leq i-1}^{-1})_{lm}\right],
  \label{infoold}
\end{eqnarray}
where $\bssC_{\leq i}$ is the upper-left square submatrix of $\bssC$
with indices only through $i$.  Figure \ref{infodefs} shows the
difference this makes in the cumulative information in $\ln A$, using
the same cosmology as in Paper I.  The difference is small, but could
be larger for a parameter with derivative terms farther from unity
than $\ln A$.  We also show the measurements from simulations (RH06),
calculated using Eqn.\ (\ref{inforange}).

\infodefs

\subsection{Covariance matrix construction}
\label{covmatcon}
We use the same procedure for the matter power spectrum
covariance matrix as in Paper I.  The covariance of the power spectrum
in a survey of volume $V$ is the sum of a Gaussian term, which depends
on the square of the power spectrum itself, and a term involving the
(hereafter, implicitly non-linear) trispectrum \citep[SZH]{hrs,szh};

\begin{equation}
C_{ij} = \frac{1}{V}\left[\frac{(2\pi)^3}{V_{s,i}}2P(k_i)^2\delta_{ij}
+ T_{ij}\right],
\label{cijdef}
\end{equation}
where $V_{s,i}$ is the volume of shell $i$ in Fourier space
(proportional to $k_i^3$ for logarithmically spaced bins), and $T_{ij}$
is the trispectrum averaged over shells $i$ and $j$;
\begin{equation}
T_{ij} \equiv \int_{s,i}\int_{s,j}T(\bk_i,-\bk_i,\bk_j,-\bk_j)\frac{d^3 \bk_i}{V_{s,i}}\frac{d^3\bk_j}{V_{s,j}}.
\label{angav}
\end{equation}

We use the halo-model formalism \citep{cs} to get the non-linear
matter power spectrum and trispectrum.  In the halo model, the
universe is assumed to consist of virialized haloes distributed
according to leading-order perturbation theory (the first-order,
linear, power spectrum, the second-order bispectrum, and the
third-order trispectrum).  In the halo model, the power spectrum is
the sum of one- and two-halo terms, and the trispectrum is the sum of
{one-}, {two-}, {three-}, and four-halo terms \citep[][CH]{coorayphd,
chu}.  For example, the power spectrum is
\begin{eqnarray}
P(k) & = & P^{\rm 1h}(k) + P^{\rm 2h}(k)\\
     & = & M^0_2(k,k) + P^{\rm lin}(k)[M^1_1(k)]^2,
\label{p1h2h}
\end{eqnarray}
where $M^\beta_\mu$ are integrals over the halo mass function;
\begin{eqnarray}
M^\beta_\mu(k_1,\ldots,k_\mu) & \equiv & \int\int \left(\frac{m}{\bar{\rho}}\right)^\mu b_\beta(m)n(m,c)\nonumber\\
& &\times u(k_1,m,c)\cdots u(k_\mu,m,c)\,dc\,dm.
\label{ibetamu}
\end{eqnarray}
Here, $\bar{\rho}$ is the mean matter density, $m$ is the halo mass,
$c$ is the halo concentration in the NFW profile \citep{nfw96},
$b_\beta(m)$ is the $\beta$-order halo bias \citep{mjw,sshj}, and
$u(k,m,c)$ is the halo profile in Fourier space, normalized to unity
at $k=0$.  In Eqn.\ (\ref{p1h2h}), we assume that the power spectrum
$P^{\rm hh}$ of a set of haloes is a uniformly biased linear power
spectrum $P^{\rm lin}$, even though this seems not to be quite true
\citep[][SSS]{sss}.

For the covariance matrix, we use use the same halo-model inputs as
did CH, except that we use a \citet{st} halo mass function.  The
baseline cosmology is a flat concordance model, specifically that
found recently by the WMAP team \citep{wmap3}, except that $n=1$,
i.e.\ $(\Omega_m h^2 = 0.127,\, \Omega_bh^2 = 0.0223,\, n = 1,\,
\sigma_8=0.74)$.  For the input linear power spectrum, we use the
\CAMB\footnote{See \url{http://camb.info/}.}  code.

\subsection{Parameter derivatives}
\dlogs

The other major ingredients in our analysis are the derivatives of
$\ln P_i$ with respect to parameters of interest.  In Paper I, we
varied $\ln A$ by small amounts, and thus calculated $\partial\ln
P_i/\partial\ln A$ numerically.  In this Letter, we do the same with
the parameters $(\ln A,\, n,\, \ln h,\, \ln f_b,\, \ln \Omega_mh^2)$.
Here, $A$ and $n$ are the scalar amplitude and tilt of the power
spectrum, using the default \CAMB\ pivot point, at $0.05\, {\rm
Mpc}^{-1}$.  The parametrized Hubble constant $h = H_0/(100\,\rm{km\
sec^{-1}\ Mpc^{-1}})$, and $f_b$ is the baryon fraction
$\Omega_b/\Omega_m$, where $\Omega_b$ is the baryon density, and
$\Omega_m$ is the sum of $\Omega_b$ and the dark-matter density,
$\Omega_c$.  We assume a flat Universe throughout.

We tried two different methods to calculate the parameter derivatives:
using our halo model code; and, using the \halofit\, (implemented in
\CAMB) fitting formula developed by \citet{sea}.  Each method has its
advantages: using our code would be self-consistent, but \halofit\,
has been extensively tested against $N$-body simulations.  We expected
the results to be similar, though, since \halofit\, was developed in
the spirit of halo models, having both quasi-linear and self-halo
terms.

Figure \ref{dlogs} is a comparison of the derivative terms from the
two methods.  While they give similar results on linear scales, there
are qualitative differences on non-linear scales.  In general,
\halofit\, predicts greater derivative terms than our code.  We also
show $\partial\ln P(k)/\partial\ln A$ as measured from 400
128-particle PM $N$-body simulations of box size $256\ihMpc$ run by
RH05 (using a slightly different cosmology).  Although both methods
are roughly consistent with the simulation measurement, the continued
decrease on smaller scales which occurs in our code is somewhat more
plausible than the upturn seen with \halofit.  The methods differ most
dramatically for $\ln h$; our halo model code predicts a tiny
variation with $\ln h$ on small scales, while \halofit\, predicts a
variation comparable to other cosmological parameters.

The two models give different predictions on small scales because of
different ways the one-halo terms $P^{1\rm h}$ (dominant on small
scales) are defined.  In the halo model, $P^{1\rm h}$ depends only on
the abundances and concentrations of haloes of different masses; these
depend on the variance in the linear density field smoothed with a
top-hat filter of radius $r$, $\sigma^2(r)$.  Keeping all of our other
parameters fixed, changing $h$ merely shifts $\ln P^{\rm lin}(k)$
horizontally and vertically.  For this cosmological model, the shift
is such that $P^{\rm lin}(k)$ does not change if the local slope ${\rm
d}\ln P^{\rm lin}(k)/{\rm d}\ln k$ reaches a value $\approx -2$.  This
happens at small scales, so $\sigma^2(r)$ does not change appreciably
for small $r$.  Thus, the abundance and concentration of small haloes
varies only slightly with $\ln h$, and $P^{1\rm h}$ hardly changes on
small scales.  On the other hand, the \halofit\ power spectrum does
vary with $\ln h$ on small scales, since it explicitly depends on
$\Omega_m$.  (In both cases, we hold $\Omega_mh^2$ constant.)

\section{Results}
The parameter Fisher matrix $F_{\alpha\beta}$ gives predictions of
statistical error bars and error ellipsoids, assuming that the
likelihood functions are Gaussian. This assumption does not precisely
hold, but is adequate to look for trends.  Holding all other
parameters fixed, the variance in a parameter $\alpha$ is
$1/F_{\alpha\alpha}$, while the variance marginalized over other
parameters is $(\bssF^{-1})_{\alpha\alpha}$, where $\bssF$ is the
Fisher matrix including $\alpha$ and the other parameters.

Figure \ref{infomarg}a shows how constraints on our chosen parameters
change with the smallest scale (largest wavenumber) used in the
analysis.  We use a survey volume of $256\hMpc$, and a fixed lowest
wavenumber, $0.02\ihMpc$ $\approx 2\pi/(256\hMpc)$.  In Fig.\
\ref{infomarg}, instead of the rather abstract quantity of
information, we show probably more familiar error-bar half-widths.

\infomarg

Diagonal plots show error-bar half-widths, both unmarginalized,
$1/\sqrt{F_{\alpha\alpha}}$ (black), and marginalized over all four
other parameters, $\sqrt{(\bssF^{-1})_{\alpha\alpha}}$ (green).  The
solid and dashed curves use the halo-model and \halofit\, derivative
terms, respectively.  For example, the information about $\ln A$ as
plotted in Fig.\ \ref{infodefs} appears (to the $-1/2$ power) in the
solid black curve in the upper-left plot.

Off-diagonal plots show correlation coefficients $R_{\alpha\beta}
\equiv
(\bssF^{-1})_{\alpha\beta}/\sqrt{(\bssF^{-1})_{\alpha\alpha}(\bssF^{-1})_{\beta\beta}}$,
in the marginalized parameter covariance matrix $\bssF^{-1}$
containing all five parameters.  If $R_{\alpha\beta} \approx \pm 1$,
then an error ellipse is squashed along a diagonal line; if
$R_{\alpha\beta} = 0$, then the ellipse is circular.

The main conclusions of this Letter come from Fig.\ \ref{infomarg}a.
The cumulative information in a parameter varied alone generally has
the characteristics found in RH05 and Paper I: there is a plateau on
translinear scales, followed by a rise on fully non-linear scales.
This rise appears as a drop in Fig.\ \ref{infomarg}a, which shows the
information to the power $-1/2$.  Unfortunately, this small-scale
error-bar tightening is quite degenerate among parameters.  The green
curves on the diagonal display this clearly; with the possible
exception of the tilt, marginalized error-bar half-widths level off in
the translinear regime, and never significantly decrease as smaller
scales are included in the analysis.  This degeneracy is also visible
in the correlation coefficients, many of which are nearly $\pm 1$ on
small scales.  Small scales provide an increased lever arm for
constraining the tilt on small scales, but the marginalized error bars
in the tilt only tighten significantly if using our code's derivative
terms, not those from \halofit.

Why is the small-scale rise in information so degenerate among
parameters?  As discussed above, $P^{\rm 1h}$ is entirely determined
by the abundances and concentrations of haloes, which depend on
integrals (over top-hat window functions) of the linear power
spectrum.  Altering any single parameter will indeed change these
integrals, but this change will generally be close to a monolithic
shift up or down in $P^{\rm 1h}$.  Thus, changing any parameter will
have a similar effect on the small-scale power spectrum.

Our previous plots have shown the cumulative information up to a
wavenumber $k_{\rm max}$, holding $k_{\rm min}$ and the volume fixed.
In Fig.\ \ref{infomarg}b, we crudely explore the effects of survey
size.  We show unmarginalized error-bar half-widths, computed using
derivatives from our code (not \halofit), changing the volume of the
survey, but not the range of scales measured.  For a box size $b$, we
assume that $P(k)$ can be measured from $k_{\rm min} = 2\pi/b$ to
$k_{\rm max} = 10 k_{\rm min}$.  Here, the translinear plateau takes
the form of a steep ramp upward at $k_{\rm max}\approx 1\ihMpc$.

\subsection{Looking in rural areas of the Universe}
In Paper I, we argued that the translinear information plateau in
$P(k)$ is caused by cosmic variance in the number of the largest
haloes in a given volume, since on translinear scales, large haloes
dominate $P(k)$.  A potential way around this problem is to model the
contribution of the largest haloes to the power spectrum.  Another way
could be to remove the `noise' of the largest haloes from the
analysis.  Even beyond the cosmic-variance argument, it is plausible
that cosmological information is especially obscured in large haloes,
in advanced stages of non-linear collapse.  This is not a new idea;
for example, much cosmological information seems to lie in the
low-overdensity Lyman-alpha forest \citep[e.g.][]{croft,gh}.

To investigate the potential of cutting out large haloes from the
analysis, we truncate the halo-model integrals over the halo mass
function at various upper mass cut-offs.  For this calculation, we use
only the dominant 1h and 2h terms in the trispectrum; we explain why
in the rest of the paragraph.  In this Letter, we assume that the
$P^{\rm hh}$ and $T^{\rm hh}$ are given by
leading-order-perturbation-theory (LOPT).  This is not quite the case
(SSS), but more accurate estimates of $P^{\rm hh}$ and $T^{\rm hh}$
are not currently known.  The LOPT power spectrum $P^{\rm lin}$ is
first-order, but the LOPT trispectrum $\to3$ is third-order; as SZH
point out, using both of them together is inconsistent.  Indeed, we
find that doing so in Eq.\ \ref{cijdef} gives off-diagonal entries
exceeding unity in the correlation matrix, violating the Schwarz
inequality.  These unruly entries are in the translinear regime, where
$\to3$ may still accurately trace the non-linear trispectrum, but
$P^{\rm lin}$ does not trace the non-linear power spectrum.  Using the
full halo mass spectrum in the halo model, this inaccuracy does not
matter substantially, since where $P^{\rm lin}$ and $\to3$ may not
accurately trace $P^{\rm hh}$ and $T^{\rm hh}$ (in the non-linear
regime), the terms in the trispectrum which involve them are buried
under other terms.  So, we use all terms in the halo-model trispectrum
when using the full halo mass function.  On the other hand, if large
haloes are missing from the mass spectrum, the raw $P^{\rm hh}$ and
$T^{\rm hh}$ are exposed on translinear scales, and the inconsistency
of using $P^{\rm lin}$ and $\to3$ together matters.  One way to keep
the order of perturbation theory consistent would be to calculate all
functions to third order, a laborious task that might not give a more
accurate result.  Instead, we assume, as before, that $P^{\rm hh} =
P^{\rm lin}$, but exclude terms involving functions of higher than
first order in the halo-model trispectrum (the 3h and 4h terms).

Figure \ref{masscut} shows that cutting out the largest haloes from
the mass function for all quantities indeed shifts the information
plateau to smaller scales, giving tighter error bars.  We also show
the results if all terms are included in the trispectrum and no mass
cut-off is made; this shows that the 1h and 2h terms do dominate the
halo-model trispectrum in this case.

\masscut

There are many practical problems which would make it difficult to
constrain cosmological parameters by removing large haloes from the
analysis.  Halo masses are difficult to measure, and measuring all
halo masses in a survey seems nearly impossible.  But perhaps
something as simple as the number of galaxies in a halo is well-enough
correlated with halo mass to make a dent in the translinear plateau.
Other problems could include inadequate knowledge of halo power
spectra on non-linear scales, and the effects on a survey mask from
excising haloes.  Still, the information gains with mass cut-offs are
dramatic enough that it seems to be worth exploring how to get around
these issues.

\section{Conclusion}
In the context of the halo model, we find that the matter power
spectrum is rather disappointing for cosmological parameter estimation
on scales smaller than linear.  On translinear scales ($k\sim
1-10\ihMpc$), there is a high degree of intrinsic (co)variance in the
matter power spectrum caused by cosmic-variance fluctuations in the
number of large haloes, which suppress the information in any
parameter of current interest on those scales.  There is information
on even smaller scales if each parameter is varied alone, but we find
that this information is quite degenerate among various cosmological
parameters.  This is because changing any parameter affects the
small-scale matter power spectrum in a similar way, close to a uniform
shift up or down in the one-halo term.  There could be more
independent information in the tilt on small scales, but one method we
used (involving \halofit) to calculate the non-linear power spectrum
predicts that the information in the tilt is also somewhat degenerate.

Our results suggest that useful cosmological information is scant
below linear scales in the full matter power spectrum, but this is
probably not the case for all large-scale structure statistics.  For
example, we show that in the halo model, the power spectrum of matter
outside of large haloes has an information plateau on smaller scales
than does the full matter power spectrum, allowing tighter constraints
on cosmological parameters.  There are practical problems with this
specific approach, but it raises hopes that there are ways to
circumvent the covariance and degeneracies among cosmological
parameters which haloes introduce on non-linear scales.  In addition,
the smallest scales of the galaxy power spectrum contain information
about galaxy formation.  Despite the apparent difficulties we have
found, we remain confident that worthwhile information exists on
non-linear scales in the spatial distribution of certain sets of
galaxies, matter, or even haloes themselves.

\section*{Acknowledgments}
We thank Andrew Hamilton, Chris Rimes, Andrew Liddle, Adrian Pope, and
Gang Chen for helpful discussions.  We are grateful for support from
NASA grants AISR NAG5-11996 and ATP NAG5-12101, and NSF grants
AST-0206243, AST-0434413 and ITR 1120201-128440.

\end{document}